\documentclass[3p,twocolumn,times]{elsarticle}
\usepackage{amsmath}
\usepackage{amssymb}
\usepackage{amsthm}
\usepackage{graphics}
\usepackage{epsfig}
\usepackage[colorlinks=true,citecolor=blue,linkcolor=red,linktocpage=true,pagebackref=false]{hyperref}
\usepackage{soul} 
\usepackage[usenames, dvipsnames]{color}
\usepackage{braket}
\usepackage{changes}
\begin{document}

%%%%%%%%%
\begin{frontmatter}

\title{
Single-electron emission from degenerate quantum levels
}

\author{Michael Moskalets}
\ead{michael.moskalets@gmail.com}
\address{Department of Metal and Semiconductor Physics, NTU ``Kharkiv Polytechnic Institute", 61002 Kharkiv, Ukraine}

\date\today
\begin{abstract}
Single-electron sources on-demand are requisite for a promising fully electronic  platform for solid-state quantum information processing. 
Most of the experimentally realized sources use the fact that only one electron can be taken from a singly occupied quantum level. 
Here I take the next step and discuss emission from orbitally degenerate quantum levels that arise, for example, in quantum dots with a nontrivial ring topology. 
I show that degeneracy provides additional flexibility for single- and two-electron emission.  
Indeed, the small Aharonov-Bohm flux, which slightly lifts the degeneracy, is a powerful tool for changing the relative width of the emitted wave packets over  a wide range.
In a ring with one lead, even electrons emitted from completely degenerate levels  can be separated in time if the driving potential changes at the appropriate rate.  
In a ring with two leads, electrons can be emitted to different leads if one of the leads has a bound state with Fermi energy at its end. 
\end{abstract}


\end{frontmatter}

%\tableofcontents
%%%%%%%%%

%%%%%%%%%%%%%
%%%%%%%%%%%%%
\section{Introduction}
\label{sec1}

The implementation of single-particle sources that inject electrons  on demand into the mesoscopic conductor \cite{Blumenthal:2007ho,Feve:2007jx,Splettstoesser:2017jd,Bauerle:2018ct}  is a milestone in the development of a fully electronic platform for quantum information purposes. 
Moreover, by analogy with quantum optics,\cite{Walls2008} of particular interest is the injection of a pair of electrons, one signal and one idle. 
Injection of more than one electron at a time has been experimentally achieved using a Lorentzian voltage pulse \cite{Dubois:2013ul,Glattli:2016tr,Bisognin:2020dk} or a dynamic quantum dot \cite{Fricke:2013cc,Fletcher:2012te,dHollosy:2015ez,Waldie:2015hy}. 
As suggested in Refs.~\cite{Splettstoesser:2008gc,Moskalets:2019us}, two electrons (holes) can be injected into the same wave guide using two single-electron sources in series. 

\begin{figure}[t]
\centering
\resizebox{0.99\columnwidth}{!}{\includegraphics{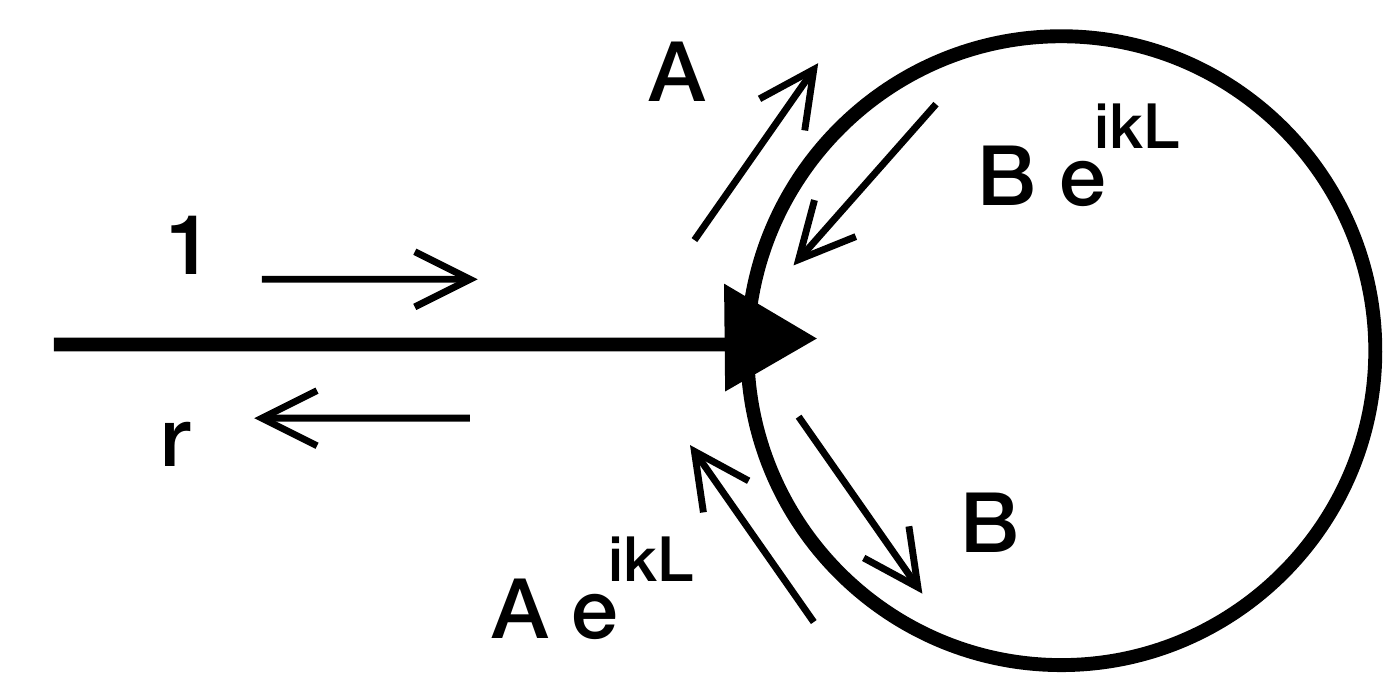} }
\caption{
One-dimensional ring of length $L$, a paradigmatic example of a system with orbitally degenerate quantum levels, is attached to the semi-infinite lead. 
The quantum-mechanical amplitudes $1$, $r$, $A$, and $B$ describe the scattering of a unit electron flux incoming from the lead in the tri-junction.  
Arrows indicate incoming and outgoing electron waves. 
The poles of the reflection coefficient $r$ as a function of the energy,  which depends on the wave number $k$, provides information on the levels in the ring. 
}
\label{fig1}
\end{figure}

Here I discuss another possibility, namely the emission of a pair of electrons from orbitally degenerate quantum levels, see Fig.~\ref{fig1}. 
The advantage of this approach is the high tunable separation of particles by  energy and emission time. 
Moreover, if the degenerate levels are coupled to two leads, electrons can be spatially separated and injected into different leads. 
This can be achieved if the leads are characterized by different topological quantum numbers, the parity of the number of bound states at Fermi energy at their ends.\cite{Fulga:2011dv} 
In this case, each level is coupled to only one lead and, therefore, each electron is injected into a separate lead. 

%%%%%%%%%%%%%
%%%%%%%%%%%%%

To date, most quantum-coherent electron circuits \cite{Bocquillon:2012if,Bocquillon:2013dp,Bocquillon:2013fp,Jullien:2014ii,Kumada:2015ch,Kataoka:2015ta,{Freulon:2015jo},Marguerite:2016jt} have been built using quantum Hall states \cite{Klitzing:1980kwa,Halperin:1982tb,Buttiker:1988vk,Buttiker:2009bg} as waveguides, since they provide a fairly long  decoherence length.\cite{{Duprez:2019wy},Clark:2020ws} 
In addition, the magnetic field increases the stability of the single-electron injection regime.\cite{Wright:2008kc,Kaestner:2009cm} 
Surprisingly, however, levitons, single-electron excitations in a ballistic conductor without a magnetic field, exhibit ideal antibunching,\cite{Dubois:2013ul,Glattli:2016tr,Glattli:2016wl} while single electrons propagating in quantum Hall states exhibit antibunching, which is not so ideal \cite{{Bocquillon:2013dp},Freulon:2015jo,Marguerite:2016jt,{Marguerite:2016ur}}. 
Therefore, systems without a magnetic field, in which degenerate quantum levels are quite expected, retain the potential for manipulating single-electron  states.

The paper is organized as follows: 
In Sec. \ref{sec2}, I consider a one-dimensional ring  connected to a semi-infinite lead. 
The time-dependent potential $U(t)$ is used to change the position of  orbitally degenerate levels in the ring relative to the Fermi energy of the electrons in the wire. 
When degenerate levels cross the Fermi energy, only one electron gets injected. 
The second electron is decoupled from the lead. 
To remove the degeneracy, I add a small Aharonov-Bohm flux \cite{Aharonov:1959wh}, $ \Phi \to 0$,  piercing the ring. 
Now the second electron can also be injected. 
It is important to note that the width of its wave-packet and therefore its energy  is highly depend on the magnetic flux and differs significantly from the width of  the wave packet of another emitted electron.
Moreover, by changing the rate of the potential $U(t)$, one can achieve a regime in which one electron is injected adiabatically, and the other in injected non-adiabatically, with a  significant delay.
In Sec. \ref{sec3}, I consider a one-dimensional ring connected to two semi-infinite leads, and I demonstarte the conditions under which two electrons are injected  into different leads.
I conclude in Sec. \ref{concl}.

%%%%%%%%%%%%%
%%%%%%%%%%%%%
\section{Ring with one lead}
\label{sec2}

The prototype of a system with discreet orbitally degenerate levels is a one-dimensional  ring of ballistically moving electrons. 
When the lead is attached to it, see Fig.~\ref{fig1}, the levels get shifted and widened. 
However, if the lead is coupled symmetrically to the left- and right-moving electrons in the ring, one of the levels does not feel the presence of the lead and is thus effectively decoupled from it, see, e.g. Refs.~\cite{Guevara:2003kf,Voo:2005hg,Vargiamidis:2006de,Hsu:2016dv,{Donarini:2018hy},Ho:2019br,Lavagna:2020wc,Aksenov:2020tq}.  
This is similar to the effect of a delta potential-like impurity on the spectrum of electrons in an isolated ring. 
The impurity affects the state that is symmetric with respect to the position of the impurity. 
And the other, anti-symmetrized state, does not depend on the presence of an  impurity at all.  
The case of a symmetric coupling of a ring to a lead is of my  interest here. 

To cause electrons to transfer from the ring to the lead, we use the nearby gate to apply the time-dependent electric potential $U(t)$ to the ring. 
As occupied doubly degenerate levels in a ring raise above the Fermi level $ \mu$ of electrons in a lead, one electron is transferred from the ring to the lead. 
Another electron stays in the ring. 
To assist the remaining electron to transfer into a lead, we need to remove degeneracy, say, by piercing a ring with an Aharonov-Bohm flux $ \Phi$. 
My aim here is to investigate how the injection of the second electron gets suppressed at $ \Phi \to 0$. 

I neglect the electron-electron interaction, which can be effectively screened out with the help of a nearby metallic gate. \cite{Feve:2007jx} 
For simplicity, I also neglect the electron spin, which can be easily incorporated into the formalism presented below. 
Additional degrees of freedom, such as spin or more exotic ones, such as  valley, make the behavior richer, but they do not compromise the main idea of this paper, according to which degeneration provides extra flexibility in the process of a single-electron emission. 

%%%%%%%%%%%%%%%
%%%%%%%%%%%%%%%
%%%%%%%%%%%%%%%
\subsection{Excess correlation function}

To obtain information about the quantum state transferred from the ring to the  lead, I  use an excess first-order correlation function  $G ^{(1)}$ \cite{Grenier:2011js,Grenier:2011dv,Haack:2013ch,Moskalets:2013dl} calculated for  electrons in the lead scattered from the ring. 

Let $\hat \Psi_{out}(t,x_{0})$ be the second quantization operator for electrons in the lead moving out of the ring evaluated at time $t$ and at a fixed position $x_{0}$. 
Then a two-time correlation function is defined as follows,  ${\cal G} ^{(1)}\left( t_{1}; t_{2} \right) = \left\langle \hat \Psi_{out}^{\dag}(t_{1}) \hat \Psi_{out}(t_{2})  \right\rangle$, where the angle brackets stand for the quantum statistical average over the equilibrium state of electrons in the lead before encountering a ring. 
For brevity, I drop $x_{0}$ from the arguments.

The excess correlation function is the difference of correlation functions calculated with a time-dependent potential $U(t)$ being switched on and off,
 $G ^{(1)} =  {\cal G} ^{(1)}_{on} - {\cal G} ^{(1)}_{off}$. 
For {\it slowly} varying $U(t)$ and at zero temperature, the excess correlation function is expressed in terms of the reflection coefficient $r$ evaluated at the Fermi energy $ \mu$, see Fig.~\ref{fig1}, \cite{{Haack:2011em},Moskalets:2015ub}

\begin{eqnarray}
v_{ \mu} G ^{(1)}\left( t_{1}; t_{2} \right) &=& 
 \frac{ e^{i  \frac{\mu }{ \hbar }\left( t_{1} - t_{2} \right) } }{ 2 \pi i  }
\frac{ r^{*}( t_{1}) r( t_{2}) - 1  }{ t_{1} - t_{2}  },  
\label{01}
\end{eqnarray}
\ \\ \noindent 
where $v_{ \mu}$ is the velocity of electrons with Fermi energy in the lead, and the star, $^{*}$, stands for the complex conjugation. 
 
If $U(t)$ causes one electron to be removed from the ring and to be appeared with the wave function $ \Psi(t)$ in the lead, one can anticipate that  $v_{ \mu} G ^{(1)}\left( t_{1}; t_{2} \right) = \Psi^{*}\left( t_{1} \right) \Psi\left( t_{2} \right)$. \cite{Grenier:2011js} 
If two electrons are transferred, then $v_{ \mu} G ^{(1)}\left( t_{1}; t_{2} \right) = \sum_{j=1}^{2}\Psi_{j}^{*}\left( t_{1} \right) \Psi_{j}\left( t_{2} \right)$.

%%%%%%%%%%%
%%%%%%%%%%%
%%%%%%%%%%%
\subsection{Ring-lead coupling}
 
I suppose that the tri-junction in Fig.~\ref{fig1} (black triangle) is characterized by an energy independent one-parameter scattering matrix, \cite{{Buttiker:1984vn},Buttiker:1985vt}

\begin{eqnarray}
\hat S &=& \begin{pmatrix} 
- \gamma & 
\epsilon  & 
\epsilon   \\
\epsilon  & 
\frac{\gamma -1 }{2 }  & 
\frac{\gamma +1 }{2 } \\
\epsilon & 
\frac{\gamma +1 }{2 }  & 
\frac{\gamma -1 }{2 }  \\
\end{pmatrix}.
\label{02}
\end{eqnarray}
\ \\ \noindent 
Here $ \gamma = \sqrt{1 - 2 \epsilon^{2}}$ to ensure unitarity, $\hat S^{\dag} \hat S = \hat I$, where $\hat I$ is a unit $3 \times 3$ matrix.

The case $ \epsilon = 0$ describes the lead, the channel $1$, completely decoupled from the ballistic now ring, the channels $2$ and $3$. 
The lead is symmetrically coupled to the right- and left-moving electrons in the ring, the channels $2$ and $3$, respectively, $S_{12} = S_{13} = \epsilon$. 
Below I am interested in the case of $ \epsilon \ll 1$, when there are well defined quantum levels in the ring. 
Note that this regime does not imply $ \epsilon \to 0$, when the Coulomb interaction effects can play a role. \cite{{Gabelli:2006eg},Feve:2016ee}

The scattering matrix $\hat S$ relates the amplitudes of plane waves with wave number $k$ incoming to and outgoing from the tri-junction as follows, see Fig.~\ref{fig1},

\begin{eqnarray}
\begin{pmatrix} 
r\\
A \\
B
\end{pmatrix} = 
\hat S 
\begin{pmatrix} 
1\\
B e^{i k L} \\
A e^{i k L}
\end{pmatrix} ,
\label{03}
\end{eqnarray}
\ \\ \noindent
where $L$ is is the length of the ring. 

Solving the three linear equations above, we find the reflection coefficient $r$. 
At $ \epsilon \ll 1$ and close to resonances, $kL - 2 \pi n \ll 1$ with $n$ integer, the reflection coefficient reads,

\begin{eqnarray}
r_{n} &=& -  \frac{  \left( kL - 2 \pi n \right)  - i  T  }{  \left( kL - 2 \pi n \right)  + i  T } .
\label{04}
\end{eqnarray}
\ \\ \noindent
where $T = \epsilon^{2}$ is the transmission probability. 

As a function of $k$, the reflection coefficient shows a set of single poles, $k^{pole}_{n}L = 2 \pi n - i T$, confirming the fact that only one in each pair of  degenerate levels in the ring is coupled to the lead.

%%%%%%%%%%%
%%%%%%%%%%%
%%%%%%%%%%%
\subsection{Single-electron emission}

When a time-dependent electric potential $U(t)$ is uniformly applied  to the ring, in addition to the kinematic phase $kL$, an electron with energy $E$  acquires the dynamic phase $ \varphi(t,E) = \frac{ - e }{ \hbar } \int _{t - \tau(E)}^{ t} dt ^{\prime} U(t ^{\prime})$, where $-e$ is the electron charge, $ \tau(E) = L/v(E)$ is the time of a single revolution, $v(E)$ is the velocity of an electron with energy $E$. \cite{Moskalets:2008fz}
As a result, the position of resonances in the ring changes with time. 
I am interested in the case when only one resonance, that is, two  degenerate levels, say with $n=n_{ \mu}$, crosses the Fermi level of electrons in the lead. 
The corresponding reflection coefficient I denote as $r(t)$. 

With the exception of Sec.~\ref{semi}, I restrict myself to the adiabatic emission regime when $U(t)$ changes slow enough. 
In this case, the dynamic phase  becomes, $\varphi(t)  = e U(t) \tau/ \hbar$. 
Note that I drop the energy argument for quantities evaluated at the Fermi energy, $E = \mu$. 

To calculate $r(t)$, I replace $kL \to k_{ \mu}L + \varphi(t)$ in Eq.~(\ref{03}).
Close to the time $t_{0}$, when the resonance in question hits the Fermi level, $k_{ \mu} L + \varphi(t_{0}) = 2 \pi n_{F}$, I linearize the dynamic phase, $ \varphi(t) \approx \varphi(t_{0}) - s (t - t_{0})$, where $s = \frac{ e \tau }{ \hbar } \left. \frac{  d U }{  d t }  \right |_{t = t_{0}}$.  
Without loss of generality, hereinafter I set $t_{0} = 0$.
Then we arrive at the reflection coefficient, 

\begin{eqnarray}
r(t) &=& - \frac{ t  + i  \Gamma _{\epsilon} }{ t  - i  \Gamma _{\epsilon} } ,
\label{05}
\end{eqnarray}
\ \\ \noindent
where $  \Gamma _{\epsilon} = T/ s$. 

I substitute $r(t)$, Eq.~(\ref{05}), into Eq.~(\ref{01}) and find that, as we anticipated, $v_{ \mu} G ^{(1)}\left( t_{1}; t_{2} \right) =  \frac{ \Gamma _{\epsilon} }{ \left | \Gamma _{\epsilon} \right | } \Psi^{*}\left( t_{1} \right) \Psi\left( t_{2} \right)$, where \cite{Keeling:2008ft}

\begin{eqnarray}
\Psi\left( t \right) &=& \sqrt{\frac{ \left | \Gamma _{\epsilon} \right | }{ \pi  } } \frac{ e ^{- i \frac{ \mu }{ \hbar } t } }{ t -  i \Gamma _{\epsilon}  }. 
\label{06}
\end{eqnarray}
\ \\ \noindent
This wave function is normalized, $ \int _{}^{ } dt \left | \Psi \right |^{2} = 1$. 
Therefore, during crossing, one electron is emitted with the above wave function. 
$\Gamma _{ \epsilon}$ is the half-width of the corresponding wave packet. 
The energy of the emitted electron, measured from the Fermi level, is $ {\cal E}_{} =  \int _{}^{ } dt \left( \hbar {\rm Im} \left [ \frac{  \partial \Psi^{*} }{  \partial t } \Psi \right] - \mu \left | \Psi \right |^{2} \right)= \hbar/ \left( 2  \Gamma _{\epsilon} \right) $. \cite{Moskalets:2009dk,Dashti:2019ts}

Note that the same wave function \cite{Grenier:2013,glattli2016method} describes a leviton \cite{Dubois:2013ul,Jullien:2014ii}, a single-particle excitation on top of the Fermi sea, created using a voltage pulse of Lorentzian shape with an integer flux \cite{Levitov:1996,Ivanov:1997,Keeling:2006}.

The associated electric current, $I(t) = -e v_{ \mu}  G ^{(1)}\left( t;t \right)$ $=  \frac{ -e }{ 2 \pi i } r(t) \frac{  \partial r^{*}(t) }{  \partial t }$, has a Lorentzian shape, \cite{Moskalets:2016fm}

\begin{eqnarray}
I(t) &=&  \frac{  \Gamma _{\epsilon} }{ \pi} \frac{ -e }{ t^{2} +  \Gamma _{\epsilon}^{2} }.
\label{07}
\end{eqnarray}
\ \\ \noindent
Note that the sign of $  \Gamma _{\epsilon}$ defines the sign of the current pulse appeared in the lead. 
When the potential increases, $\left. \frac{  d U }{  d t }  \right |_{t = t_{0}} > 0 \Rightarrow  \Gamma _{\epsilon} > 0$, an electron is injected into the lead. 
When the potential decreases, $\left. \frac{  d U }{  d t }  \right |_{t = t_{0}} < 0 \Rightarrow  \Gamma _{\epsilon} < 0$, a hole is injected into the lead, that is, an electron is transferred from the lead to the ring.
For definiteness, below I use $  \Gamma _{\epsilon} > 0$. 

%%%%%%%%%%%
%%%%%%%%%%%
%%%%%%%%%%%
\subsection{Two-electron emission}

To facilitate two-particle emission, we lift the degeneracy, for instance, by inserting the Aharonov-Bohm flux, $ \Phi$, into the ring.    
Then the equation~(\ref{03}) becomes, 

\begin{eqnarray}
\begin{pmatrix} 
r\\
A \\
B
\end{pmatrix} = 
\hat S 
\begin{pmatrix} 
1\\
B e^{i \left [ k L +  \varphi(t) - \phi  \right]} \\
A e^{i \left [ k L +  \varphi(t) + \phi  \right]}
\end{pmatrix} ,
\label{08}
\end{eqnarray}
\ \\ \noindent
where $ \phi = 2 \pi \Phi /  \Phi_{0}$ with $ \Phi_{0} = h / e $ being the normal-metal magnetic flux quantum \cite{Buttiker:1983wd,Webb:1985tu,Chandrasekhar:1985up}. 

I am interested in the small magnetic flux limit, $ \phi \to 0$. 
Now, close to the time when the driven level crosses the Fermi level, the reflection coefficient takes the following form, 

\begin{eqnarray}
r(t) &=& -  \frac{  t ^{2} - \Gamma _{\phi}^{2} + i  t \Gamma _{ \epsilon}  }{ t ^{2} - \Gamma _{\phi}^{2} - i  t \Gamma _{ \epsilon}  }, 
\label{09}
\end{eqnarray}
\ \\ \noindent
where  $\Gamma_{ \phi} =  \phi / c$. 
Obviously, for $ \phi = 0$ we recover Eq.~(\ref{05}) for all times but $t=0$. 

Substituting $r(t)$, Eq.~(\ref{09}), into Eq.~(\ref{01}), I find that now the correlation function is represented by the sum of two single-particle contributions, $v_{ \mu} G ^{(1)}\left( t_{1}; t_{2} \right) = \sum_{j=1}^{2}\Psi_{j}^{*}\left( t_{1} \right) \Psi_{j}\left( t_{2} \right)$.  
I emphasize that  the precise equations for $ \Psi_{1}$ and $ \Psi_{2}$ are not unique, see Refs.~\cite{Dubois:2013fs,{Vanevic:2016eq},Roussel:2017hu,Vanevic:2017ci,Yin:2019bg,{Bisognin:2020dk},Yue:2020uo} for various approaches to single-particle wave functions, see also Refs.~\cite{Ryu:2016kb,Kashcheyevs:2017vc,Yamahata:2019tv,Fletcher:2019vf}. 
Unique is  $G ^{(1)}$ and the two-particle wave function, which is the Slater determinant composed of single-particle wave functions. \cite{Moskalets:2014ea}
To be specific, below I choose some specific representation, which is naturally follows from Eq.~(\ref{01}). 

The particular form of wave functions $ \Psi_{j}$ depends on the relationship between $  \Gamma _{\phi}$ and $  \Gamma _{\epsilon}$. 
For the purposes of this study, I distinguish between two regimes: large and small magnetic flux.
In the large magnetic flux regime, 
$ 2\Gamma _{\phi} > \Gamma _{\epsilon}$, the distance between the  levels ($\sim  \Gamma _{\phi}$) caused by the Aharonov-Bohm flux exceeds the width of the levels  ($\sim  \Gamma _{\epsilon}$) caused by coupling to the lead.   
In the small magnetic flux regime, 
$ \Gamma _{\epsilon} > 2\Gamma _{\phi}$, the width of the levels exceeds the distance between them.

%%%%%%%%%%%
%%%%%%%%%%%
%%%%%%%%%%%
\subsubsection{Large magnetic flux}

If two levels are well separated from each other, $ 2\Gamma _{\phi} > \Gamma _{\epsilon}$, then the emission process, when one level crosses the Fermi level, is more or less independent of the emission process, when the other level crosses the Fermi level. 
Indeed, the two wave functions, 

\begin{eqnarray}
\Psi_{1}\left(  t \right) &=&
\sqrt{ \frac{ \Gamma _{ } }{ \pi } }
\frac{ e ^{- i \frac{ \mu }{ \hbar } t } }{ t - \tau - i \Gamma_{ }  } ,
\nonumber \\
\label{10} \\
\Psi_{2}\left(  t \right) &=& 
\sqrt{ \frac{ \Gamma _{ } }{ \pi } } 
\frac{ e ^{- i \frac{ \mu }{ \hbar } t } }{ t + \tau - i \Gamma _{ }  }
e^{ 2 i \arctan\left(  \frac{ t - \tau }{ \Gamma_{ }  } \right) } , 
\nonumber 
\end{eqnarray}
\ \\ \noindent
describe the emission of single-electron wave packets of the same width $ \Gamma =  \Gamma _{\epsilon} / 2$ occuring at different times $t = \pm  \tau$, where $ \tau = \sqrt{ \Gamma_{ \phi}^{2} - \Gamma_{ } ^{2} }$. 
The energies of the emitted electrons are 
$ {\cal E}_{1} = \hbar/(2  \Gamma _{})$ and 
$ {\cal E}_{2} = {\cal E}_{1} \left( 1 + 2\left [\Gamma / \Gamma _{ \phi}  \right]^{2} \right)$, respectively. 
I emphasize, the separation into two energies is not universal and depends on representation. 
What is fixed is the total energy $ {\cal E}_{} = {\cal E}_{1} + {\cal E}_{2}$, which increases with decreasing a magnetic flux, $  \Gamma _{ \phi}$, that is, with decreasing time delay, $ \tau$.  
An increase in total energy with a decrease in the time delay between two wave packets was noted in Refs.~\cite{Moskalets:2009dk,Moskalets:2014ea}. 

With decreasing a magnetic flux, starting from $  2\Gamma _{\phi} =  \Gamma _{\epsilon}$, both wave packets are emitted at the same time $t=0$. 
Therefore, they significantly affect each other. 
We are entering a small magnetic flux regime.  

%%%%%%%%%%%
%%%%%%%%%%%
%%%%%%%%%%%
\subsubsection{Small magnetic flux}
 
When two levels are close enough to be almost degenerate, two electrons try  to escape simultaneously. 
Since electrons are fermions, this is possible if they have substantially  different energies. 
The energy of a wave packet is inversely proportional to its width. \cite{Keeling:2006}
Therefore, one can expect that at a small magnetic flux, the wave packets differ significantly in their width. 

Substituting Eq.~(\ref{09}) into Eq.~(\ref{01}), I find that for $ 2\Gamma _{\phi} < \Gamma _{\epsilon}$ the correlation function is  $v_{ \mu} G ^{(1)}\left( t_{1}; t_{2} \right) = \sum_{j=1}^{2}\Psi_{j}^{*}\left( t_{1} \right) \Psi_{j}\left( t_{2} \right)$, and the two wave functions can be chosen as follows, 

\begin{eqnarray}
\Psi_{1}\left(  t \right) &=&
\sqrt{ \frac{ \Gamma _{1 } }{ \pi } }
\frac{ e ^{- i \frac{ \mu }{ \hbar } t } }{ t  - i \Gamma_{1 }  } ,
\nonumber \\
\label{11} \\
\Psi_{2}\left(  t \right) &=& 
\sqrt{ \frac{ \Gamma _{2 } }{ \pi } } 
\frac{ e ^{- i \frac{ \mu }{ \hbar } t } }{ t  - i \Gamma _{ 2}  }
e^{ 2 i \arctan\left(  \frac{ t }{ \Gamma_{1 }  } \right) } , 
\nonumber 
\end{eqnarray}
\ \\ \noindent
where $ \Gamma_{1,2} =  \Gamma_{} \pm \sqrt{  \Gamma_{ }^{2} - \Gamma_{ \phi}^{2}  }$. 
Now the energies of the emitted particles are 
$ {\cal E}_{1} = \frac{ \hbar }{ 2 \Gamma_{1}}$ and 
$ {\cal E}_{2} = \frac{ \hbar }{ 2 \Gamma_{2}} + \frac{ 2 \hbar }{  \Gamma_{1} + \Gamma_{2}}$, respectively.

On the border between large and small magnetic flux regimes, 
$ 2  \Gamma _{\phi} =  \Gamma _{\epsilon}$, the set of wave functions in Eqs.~(\ref{10}) and (\ref{11}) are the same. 
Two electrons are emitted at the same time and the corresponding current pulses, $I_{j} \sim \left | \Psi_{j} \right |^{2}$, have the same width $  \Gamma _{\epsilon}/2$. 
However, due to the time-dependent phase factor in $ \Psi_{2}$, the wave functions $ \Psi_{1}$ and $ \Psi_{2}$ are orthogonal, $ \int _{}^{ } dt \Psi_{1} \Psi_{2}^{*} = 0$. 
Moreover, the energies of two electrons differ three times.  

An intriguing feature of the small magnetic flux regime is that the width of one of the wave packets strongly depends on the magnetic flux or any other parameter that lifts degeneracy. 
Indeed, in the limit of $ \Gamma_{ \phi} \to 0$, the width $ \Gamma_{2}$ scales quadratically with a magnetic flux, $ \Gamma_{2} \approx \Gamma_{ \phi}^{2} /  \Gamma _{\epsilon} $. 
The decrease in the width of the emitted wave packet is associated with a decrease in the effective coupling between the lead and the corresponding quantum level in the ring. 
Interesting, the width $ \Gamma_{2}$ is inversely proportional to the actual coupling between the ring and the lead, the transmission $T = \epsilon^{2}$, which is somewhat counterintuitive. 

In the limit of $ \Gamma_{ \phi} \to 0$, the width of the other wave packet is flux-independent, $ \Gamma_{1} \approx 2 \Gamma =  \Gamma _{\epsilon}$, and the wave function $ \Psi_{1}$ from Eq.~(\ref{11}) becomes the wave function $ \Psi$ from Eq.~(\ref{06}).

%%%%%%%%%%%
%%%%%%%%%%%
%%%%%%%%%%%
\subsubsection{From large to small magnetic flux}

%%%%%%%%%%%%%
\begin{figure}[t]
\centering
\resizebox{0.99\columnwidth}{!}{\includegraphics{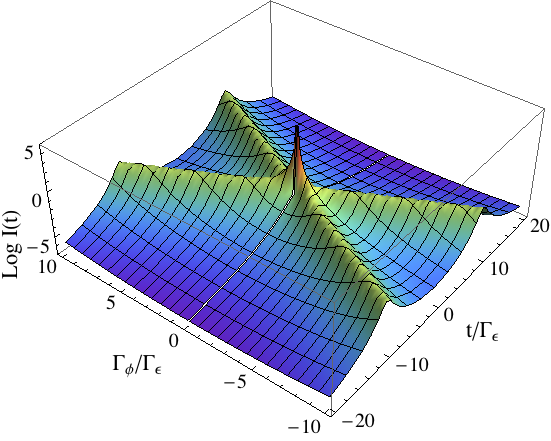} }
\caption{
Time-dependent current $I(t) = -e \left\{ \left | \Psi_{1}\right |^{2}  +  \left | \Psi_{2} \right |^{2} \right\} $ adiabatically injected from a source with two quantum levels, with splitting controlled by the parameter $ \Gamma_{ \phi}$.  
For a large splitting, $2 \left | \Gamma_{ \phi} \right | >  \Gamma _{\epsilon}$, the wave functions $ \Psi_{1}$ and $\Psi_{2}$ are given in Eq.~(\ref{10}), and  for a small splitting, $2 \left | \Gamma_{ \phi} \right | <  \Gamma _{\epsilon}$, the wave functions are given in Eq.~(\ref{11}). 
Notice the Log-scale on the vertical axis. 
The interval $\left | \Gamma_{ \phi}/  \Gamma _{\epsilon} \right | < 0.05$ is excluded.}
\label{fig2}
\end{figure}
%%%%%%%%%%%%%
In Fig.~\ref{fig2}, using a time-dependent current, $I(t) = -e v_{ \mu}  G ^{(1)}\left( t;t \right)$, as an example, I illustrate how a magnetic flux, the parameter $ \Gamma_{ \phi}$, modifies the emitted state. 

For a large separation, $ \Gamma _{\phi} / \Gamma _{\epsilon} > 0.5$, the current consists of two pulses of the same width $ \Gamma$ and height $-e/( \pi \Gamma)$. 
The pulses peak at different times, which vary with $ \Gamma_{ \phi}$, see Eq.~(\ref{10}). 

For a small separation, $ \Gamma _{\phi} / \Gamma _{\epsilon} < 0.5$, the current consists of two pulses emitted simultaneously at $t=0$, see Eq.~(\ref{11}). 
They have significantly different widths and heights in the limit of $ \Gamma_{ \phi} \to 0$. 
The ratio of the widths decreases quadratically with the magnetic flux, $ \Gamma_{2}/ \Gamma_{1} \approx \left( \Gamma_{ \phi} /  \Gamma _{\epsilon} \right)^{2}$.  
Accordingly, the ratio of the amplitudes increases quadratically. 

Note that the shape of the density profile of the emitted narrow wave packet is symmetric, Lorentzian. 
This contrasts with the asymmetric, as a rule, shape of the Fano resonance \cite{{Fano:1961ha},Miroshnichenko:2010ew} in conductance due to the same quantum level whose coupling to the leads is varied through zero.  

So, we see that no matter how small the splitting is, the two wave packets seem to be emitted. 
One of them simply becomes more and more narrow, but it always carries the  full charge of an electron. 
In the limit $ \Gamma_{ \phi} \to 0$, the density profile of the wave packet $ \Psi_{2}$ approaches the delta-function centered at $t=0$. 
This is what the adiabatic theory predicts. 
However, the adiabatic emission of such a particle requires an infinitely slow $U(t)$, and the emission process takes infinite time. 
In real experiment, this electron is emitted non-adiabatically, and its wave function $ \Psi_{2}^{na}$ differs from $ \Psi_{2}$, Eq.~(\ref{11}).

%%%%%%%%%%%
%%%%%%%%%%%
%%%%%%%%%%%
\subsection{Semi-adiabatic emission at small magnetic flux}
\label{semi}

Adiabatic emission takes place when the time during which the wave packet is formed exceeds the dwell time, the time during which an electron with a fixed energy above the Fermi level leaves the source.  \cite{Splettstoesser:2008gc,Misiorny:2017ua}
The (half-)duration of wave packets is given by $\Gamma_{1,2}$, see Eq.~(\ref{11}).
The dwell time, $ \tau_{D,j} = \hbar/ \delta_{j}$ depends on the broadening $  \delta_{j}$ of the corresponding levels, $j = 1,2$. 
This broadening can be extracted from the characteristics of the wave packet emitted adiabatically. 
Namely, the wave packet is formed while the broadened quantum level crosses the Fermi level. 
Since the level raises at a constant rapidity $c = \left. e\frac{  d U }{  d t }  \right |_{t = t_{0}}$, the parameters $ \Gamma_{j}$ and $ \delta_{j}$ are related as follows, $ \delta_{j} = c \Gamma_{j}$. 
The level rapidity $c$ is related to the phase rapidity $s$ introduced before Eq.~(\ref{05}), $c = \hbar s/ \tau$. 
Remind that $ \tau = L/ v_{ \mu}$ is the time of one revolution. 
As a result, I have, $ \tau_{D,j} = \tau/ \left( s \, \Gamma_{j} \right)$. 
In the limit $  \Gamma _{ \phi} \to 0$, two dwell times are,

\begin{eqnarray}
\tau_{D,1} &=& \frac{ \tau }{ T } ,
\label{12} \\
\tau_{D,2} &=& \tau_{D,1} \left( \frac{ T }{ \phi } \right)^{2} ,
\nonumber 
\end{eqnarray}
\ \\ \noindent 
where $T = \epsilon^{2}$ is the transmission probability, and $ \phi = 2 \pi \Phi/ \Phi_{0}$ is the phase change due to the magnetic flux $ \Phi$. 
We clearly see that $ \tau_{D,2}$ approaches infinity  with a decrease in a magnetic flux, that is, the electron occupying this level can stay longer and longer in the source.  
 
To distinguish between adiabatic and non-adiabatic emission regimes, let us introduce the non-adiabaticity parameter $ \zeta_{j} = \tau_{D,j} / \Gamma_{j}$. \cite{Keeling:2008ft} 
The adiabatic regime is realized when this parameter approaches zero.
Using Eq.~(\ref{12}), I calculate $ \zeta_{1} = s \tau/ T^{2}$ and $ \zeta_{2} = s \tau T^{2}/ \phi^{4} $. 
Since the ratio $ \zeta_{1}/ \zeta_{2} = \left( \phi/ T \right)^{4} $ decreases very quickly  with decreasing ratio $ \phi/T < 1$, there is possible a regime when one electron is emitted adiabatically, $ \zeta_{1} \ll 1$, and the other is emitted non-adiabatically, $ \zeta_{2} \gg 1$.  
I call this regime {\it semi-adiabatic}.  

In the semi-adiabatic regime, the wave function $ \Psi_{1}$ is given in Eq.~(\ref{11}). 
To determine the wave function $ \Psi_{2}^{na}$, let us proceed as follows.
I assume that the linear change in potential over time, $d U /d t = {\rm const}$, lasts long enough, not only longer than $ \Gamma_{1}$, but also longer than $ \tau_{D,2}$. 
Then we can use the solution to the problem of non-adiabatic emission from the quantum level, which raises at a constant rapidity, and write down the wave function, \cite{Keeling:2008ft}

\begin{eqnarray}
\Psi_{2}^{na}(t) &=& 
e^{ 2 i \arctan\left(  \frac{ t }{ \Gamma_{1 }  } \right) } 
\frac{ e^{ - \frac{ i }{ \hbar  } \mu  t  }  }{ \sqrt{ \pi \Gamma_{2} }  }
\int\limits_{0}^{ \infty} d x 
e^{ -  x } 
e^{  -  i x \frac{ t }{ \Gamma_{2} }   } 
e^{ i x ^{2} \zeta_{2}  } .  
\nonumber \\
\label{13}
\end{eqnarray}
\ \\ \noindent
Recall that the phase factor $e^{ 2 i \arctan\left(  \frac{ t }{ \Gamma_{1 }  } \right) } $ accounts for the effect of the emission of the other electron, see Eq.~(\ref{11}) and Ref.~\cite{Moskalets:2019us}.  

For $ \zeta_{2} \to 0$, the above equation reproduces the Lorentzian in shape $ \Psi_{2}$ from Eq.~(\ref{11}) (up to the irrelevant factor $i$) centered around $t=0$. 
In contrast, for $ \zeta_{2} \gg 1$, the above equation describes an exponentially decaying wave packet starting at $t=0$, \cite{Keeling:2008ft}

\begin{eqnarray}
\Psi_{2}^{na}(t) & \approx &  \theta(t) \frac{ C(t)  }{ \sqrt{ \tau_{D,2} } } \exp\left( - \frac{ t }{ 2 \tau_{D,2} } \right) , \quad \zeta_{2} \gg 1, 
\nonumber \\
\label{14}
\end{eqnarray}
\ \\ \noindent
where $ \theta(t)$ is the Heaviside theta function, and the phase factor $C(t) = \frac{1 + i }{ \sqrt{2} } e^{ 2 i \arctan\left(  \frac{ t }{ \Gamma_{1 }  } \right)  - i \left( \frac{ t }{2 \tau_{D,2} } \right)^{2} \zeta_{2}} $.

Since $ \zeta_{2} \sim \phi^{-4}$, when the magnetic flux $ \phi$ decreases,   the system always end up in the regime with $ \zeta_{2} \gg 1$. 
Interestingly, in this non-adiabatic regime the width of the wave packet $ \tau_{D,2}$ exceeds not only its width in the adiabatic regime, $ \tau_{D,2} / \Gamma_{2}  = \zeta_{2} \gg 1 $, but can also  exceed the width $  \Gamma _{\epsilon}$ of another electron, the one that is emitted adiabatically. 
Indeed, $ \tau_{D,2} / \Gamma_{ \epsilon}  = \zeta_{1} \left(  T /  \phi  \right)^{2} > 1 $ for $ \phi < \sqrt{ \tau s}$. 
Recall that $ \tau$ is the time of a single revolution around the ring, and $s = \frac{ e \tau }{ \hbar } \left. \frac{  d U }{  d t }  \right |_{t = t_{0}}$ is the phase rapidity, the rate of phase change due to the time-varying electric potential applied to the ring. 

%%%%%%%%%%%%%
\begin{figure}[t]
\centering
\resizebox{0.99\columnwidth}{!}{\includegraphics{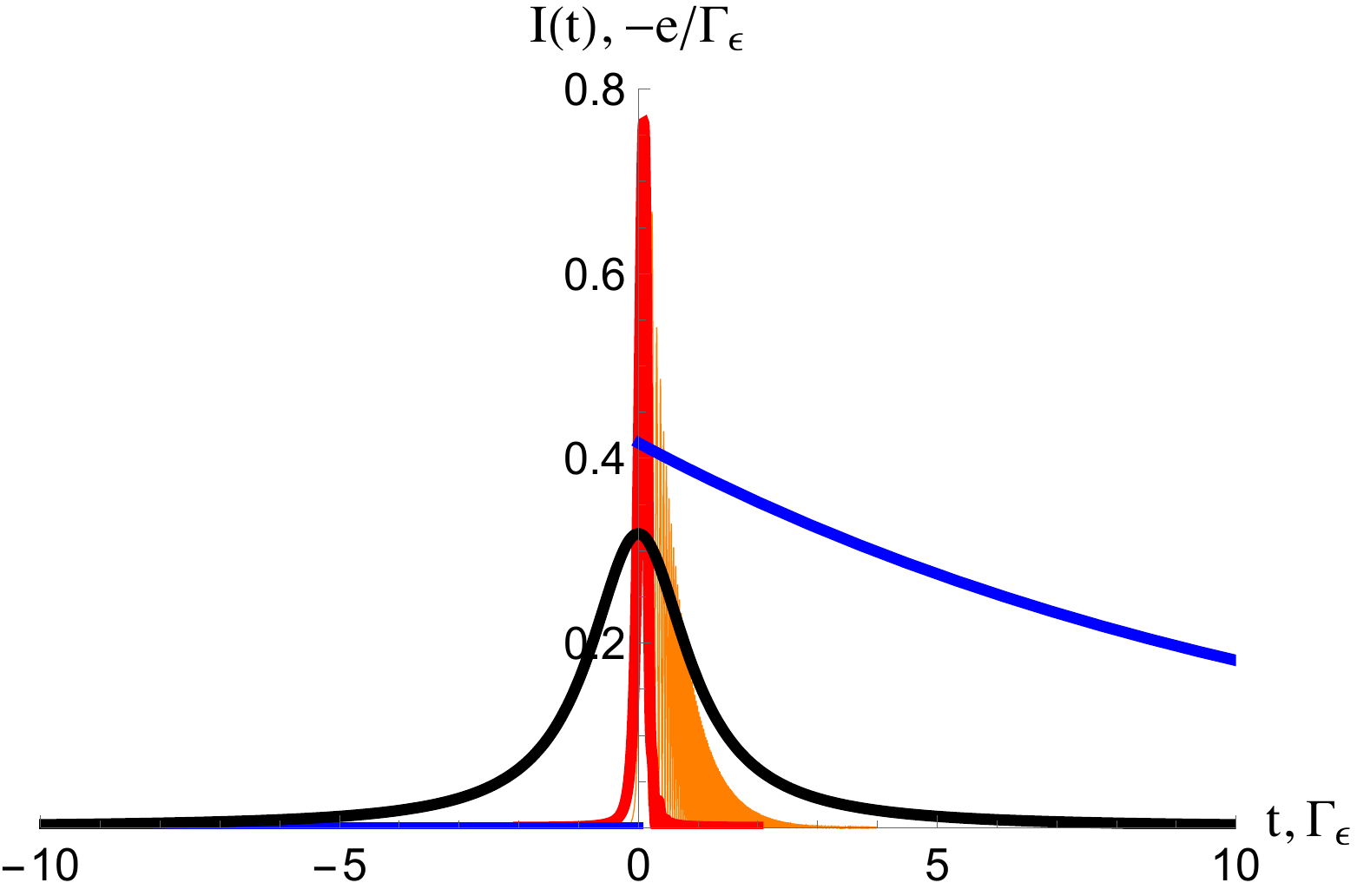} }
\caption{
Single-electron current $I(t) = -e \left | \Psi_{}\right |^{2}$ with: 
$ \Psi = \Psi_{1}$, Eq.~(\ref{11}) ({\bf black line}); 
$ \Psi = \Psi_{2}$, Eq.~(\ref{13}), for $ \phi =  0.075$, respectively, $ \zeta_{2} = 0.85$ and $ \Gamma_{2} / \Gamma_{ \epsilon}  = 0.067$, 
the value is reduced by $4$ times ({\bf red line}) and for $ \phi =  0.025$, respectively, $ \zeta_{2} = 68.4$ and $ \Gamma_{2} / \Gamma_{ \epsilon}  = 0.007$, the value is reduced by $2$ times ({\bf orange line});  
$ \Psi = \Psi_{2}^{na}$, Eq.~(\ref{14}) with $ \tau_{D,2} /  \Gamma _{\epsilon} = 12$  for $ \phi = 0.005$, respectively, $\zeta_{2} \approx 4\cdot 10^{4}$, the value is increased by $5$ times, only the envelope function is shown ({\bf blue line}). 
Other parameters: $T = 0.3$,  $s \tau = 3\cdot 10^{-4}$.
}
\label{fig3}
\end{figure}
%%%%%%%%%%%%%

To illustrate the effect of non-adiabaticity, in Fig.~\ref{fig3}  I compare the time-dependent current carried by an adiabatically emitted electron with  wave function  $\Psi_{1}$, Eq.~(\ref{11}), (black line) and the current carried by another electron with wave function  $\Psi_{2}$, Eq.~(\ref{13}), for small, $ \zeta_{2} \sim 1$ (red line), intermediate $\zeta_{2} \sim 100$ (orange line), and  large, $ \zeta_{2} \gg 1$ (blue line), non-adiabaticity parameter. 
In the latter case, I used the wave function  $\Psi_{2}^{na}$, Eq.~(\ref{14}), which allows us to reproduce the envelope function for a time-dependent current without fast oscillations visible in the orange line. 
Obviously, with increasing non-adiabaticity parameter the wave packet is  emitted at a later time, and its width increases.

Thus, in the case of symmetric coupling, Eq.~(\ref{02}), one electron is emitted from two orbitally degenerate levels, and two electrons are emitted when degeneracy is lifted, for instance, by a magnetic flux. 
The transition between two- and single-electron emission occurs in the form of an increase in the delay in the emission of the second electron.

%%%%%%%%%%%%%
%%%%%%%%%%%%%
\section{Ring with two leads}
\label{sec3}

If two leads are attached to the driven ring, see Fig.~\ref{fig4}, then in the general case an electron from the occupied level in the ring can be injected into any of the  leads. 
To be more precise, into both of them simultaneously. 
However, under specific conditions, two electrons occupying two orbitally degenerate levels are injected into different leads. 
This happens if the ring-lead couplings have different signs. 
As a result, the leads are coupled to orthogonal combinations of degenerate states. 
Keeping this effect in mind, in this section I restrict myself to the case with degenerated levels, $ \phi = 0$, and to the adiabatic regime only.

%%%%%%%%%%%%%
%%%%%%%%%%%%%
\subsection{Excess correlation matrix}

To characterize the quantum state transferred from the ring to the leads, we introduce an excess first-order correlation matrix  $\hat G ^{(1)}$ \cite{Moskalets:2016fm} calculated for electrons in the leads scattered from the ring. 

Let $\hat \Psi_{out, \alpha}(t,x_{0, \alpha})$ be the second quantization operator for electrons in the lead $ \alpha$ moving out of the ring evaluated at time $t$ and at a fixed position $x_{0, \alpha}$. 
Then a two-time correlation matrix element is defined as follows,  ${\cal G}_{ \alpha \beta} ^{(1)}\left( t_{1}; t_{2} \right) = \left\langle \hat \Psi_{out, \alpha}^{\dag}(t_{1}) \hat \Psi_{out, \beta}(t_{2})  \right\rangle$, where the angle brackets stand for the quantum statistical average over the equilibrium state of electrons in the leads before encountering a ring. 
I suppose that both leads are attached to reservoirs with the same chemical potential $ \mu$ and zero temperature. 
For brevity, I drop $x_{0, \alpha}$ and $ x_{0, \beta}$ from the arguments.
The excess correlation matrix is the difference of correlation matrices  calculated with a time-dependent potential $U(t)$ being switched on and off,
 $\hat G ^{(1)} =  \hat {\cal G} ^{(1)}_{on} -  \hat {\cal G} ^{(1)}_{off}$. 

%%%%%%%%%%%%%
\begin{figure}[t]
\centering
\resizebox{0.99\columnwidth}{!}{\includegraphics{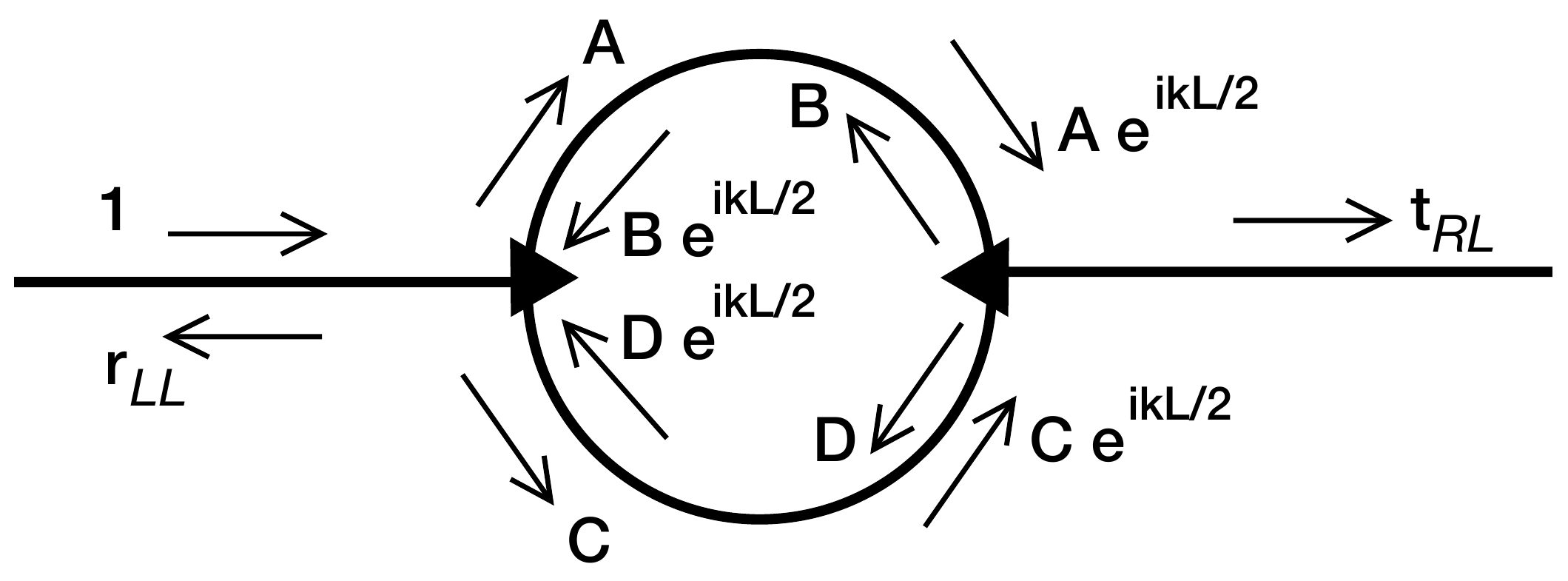} }
\caption{
A one-dimensional ring of length $L$ is symmetrically attached to two semi-infinite leads. 
A unit electron flux incoming from the left lead is reflected with an amplitude of $r_{LL}$ from the ring and transmitted through the ring with an amplitude of $t_{RL}$.  
The quantum-mechanical amplitudes $A$, $B$, $C$, and $D$ describe the propagation of an electron along a ring. 
Arrows indicate the incoming and outgoing electron waves in two tri-junctions. 
}
\label{fig4}
\end{figure}
%%%%%%%%%%%%%

For {\it slowly} varying $U(t)$ and at zero temperature, the excess correlation matrix is expressed in terms of the scattering matrix $\hat {\cal S}$ of the ring with two leads, which is evaluated at the Fermi energy $ \mu$,

\begin{eqnarray}
v_{ \mu} \hat G ^{(1)}_{}\left(  t_{1}, t_{2} \right) &=& 
\frac{ e^{i  \frac{\mu }{ \hbar }\left( t_{1} - t_{2} \right) }  }{ 2 \pi i  }
\frac{ \hat {\cal S}^{\dag}( t_{1}) \hat {\cal S} \left(  t_{2} \right) - \hat I  }{ t_{1} - t_{2}  } ,
\label{15} 
\end{eqnarray}
\noindent \\
where $\hat I$ is the unit matrix. 
 
I stress, that in the case shown in Fig.~\ref{fig4}, $\hat {\cal S}$ is a $2 \times 2$ matrix whose elements, for example, ${\cal S}_{LL} = r_{LL}$ and ${\cal S}_{RL} = t_{RL}$.    
The matrix $\hat {\cal S}$ is different from $3 \times 3$ matrices, $\hat S^{L}$ and $\hat S^{R}$, associated with tri-junctions (black triangles in Fig.~\ref{fig4}).

%%%%%%%%%%%%%
%%%%%%%%%%%%%
\subsection{Symmetrical two-lead coupling}

I will first illustrate how the excess correlation matrix $\hat G^{(1)}$ encodes information about where the quantum state is being transferred from the ring. 

For the sake of simplicity, let us suppose that both tri-junctions in Fig.~\ref{fig4} are characterized by the same scattering matrices, $\hat S^{L} = \hat S^{R}$, which are equal to $\hat S$, Eq.~(\ref{02}) with $ \epsilon \to \epsilon/ \sqrt{2}$ (to keep the level width the same). 
Then, in the limit of $ \epsilon \ll 1$, the solution of the scattering problem with a unit flux incoming from the left gives us $r_{LL} \approx \left( r -1 \right)/2$ and $t_{RL} \approx \left( r+1 \right)/2$ with $r$ being the reflection coefficient of the ring with one lead.
Solving a similar problem, but with a unit flux incoming from the right, provides $r_{RR} = r_{LL} $ and $t_{LR} = t_{RL}$. 
Thus, the scattering matrix of the ring reads,

\begin{eqnarray}
\hat {\cal S}  &=&
\frac{ 1 }{ 2 }
\begin{pmatrix} 
 r - 1  & 
r + 1  \\
r + 1  & 
r - 1
\end{pmatrix} , \quad \epsilon \ll 1,  
\label{16}
\end{eqnarray}
\noindent \\
where $r$ is given in Eqs.~(\ref{04}) and (\ref{05}).  
Recall, since the time-reversal invariance is not broken in our system, the scattering matrix is symmetric, ${\cal S}_{LR} = {\cal S}_{RL}$. 
Since both couplings are the same, the diagonal elements are equal,  ${\cal S}_{LL} = {\cal S}_{RR}$. 

Substituting the above equation in Eq.~(\ref{15}), I get an excess correlation matrix,
 
\begin{eqnarray}
\hat G ^{(1)}_{} &=&
\frac{ 1 }{ 2 }
\begin{pmatrix} 
G ^{(1)}_{}  & 
- G ^{(1)}_{}  \\
- G ^{(1)}_{}  & 
G ^{(1)}_{} 
\end{pmatrix} ,
\label{17}
\end{eqnarray}
\noindent \\
where the excess correlation function  $G ^{(1)}$ is shown in Eq.~(\ref{01}).  
Nonzero off-diagonal elements of the above matrix  $\hat G ^{(1)}$ indicate  correlations between what is injected into different leads. 
These elements are necessary to ensure that the injected state is a pure quantum state. 
That is, the matrix $\hat G ^{(1)}\left( t_{1}; t_{2} \right)$, Eq.~(\ref{17}), is idempotent provided that the function $G ^{(1)}\left( t_{1}; t_{2} \right)$ is idempotent.

With the reflection coefficient from Eq.~(\ref{05}), I find that still only one electron leaves the ring but it is now extended to two leads.  
Its wave function is a vector, $\hat \Psi\left( t_{} \right)$, with elements associated with two leads. 
Representing the correlation matrix as the direct product, 
$  \hat G ^{(1)}\left( t_{1}; t_{2} \right) = \hat \Psi^{*}\left( t_{1} \right) \otimes \hat \Psi\left( t_{2} \right)  $, I find, 

\begin{eqnarray}
\hat \Psi &=& 
\frac{1 }{ \sqrt{2} }
\begin{pmatrix} 
\Psi  \\ 
- \Psi  
\end{pmatrix} ,
\label{18}
\end{eqnarray}
\ \\ \noindent
where $ \Psi(t)$ is shown in Eq.~(\ref{06}). 
We see that in the present case, an electron  is evenly divided between the two leads. 

The situation changes dramatically if $\hat S^{L}$ and $\hat S^{R}$  no longer match. 
To consider the most spectacular case, I suppose that $\hat S^{L}$ and $\hat S^{R}$ possess different symmetry in the following sense. 
For $ \epsilon = 0$, the scattering matrix $\hat S$, Eq.~(\ref{02}), describes  a ballistic ring disconnected from a semi-infinite lead. 
The electrons in the lead acquire the phase $ \pi$ upon reflection from its end, $S_{11} = -1$, as expected for reflection from an infinite potential wall. 
However, this is not the only possibility. 
The reflection phase takes on the additional contribution $ \pi$ if there is a bound state at the endpoint of the lead.\cite{Fulga:2011dv} 
Recall, I am interested in the reflection coefficient calculated at the Fermi energy. 

In the next section, I suppose that the reflection coefficients of decoupled leads have different signs, say, $S^{L}_{11}= -1 $ and $S^{R}_{11} = 1$. 

%%%%%%%%%%%%%
%%%%%%%%%%%%%
\subsection{Anti-symmetrical two-lead coupling} 

Here I analyze the case when in Fig.~\ref{fig4} the left tri-junction is characterized by the scattering matrix $\hat S^{L} = \hat S$ with $\hat S$ given in Eq.~(\ref{02}), and the right one is characterized by the following scattering matrix, 

\begin{eqnarray}
\hat S^{R} = 
\begin{pmatrix} 
\gamma & 
-\epsilon  & 
\epsilon   \\
-\epsilon  & 
\frac{1 - \gamma  }{2 }  & 
\frac{\gamma +1 }{2 } \\
\epsilon & 
\frac{\gamma +1 }{2 }  & 
\frac{1 - \gamma}{2 }  \\
\end{pmatrix} ,
\label{19}
\end{eqnarray}
\ \\ \noindent 
where $ \gamma = \sqrt{1 - 2 \epsilon^{2}}$. 
For simplicity, the small parameter $ \epsilon \ll 1$, which determines the transmission probability, $T = \epsilon^{2}$, is the same in Eqs.~(\ref{19}) and (\ref{02}).  

The solution to the scattering problem now gives the scattering matrix of the ring as follows, 

\begin{eqnarray}
\hat {\cal S}  &=&
\begin{pmatrix} 
 r  & 
0  \\
0  & 
-r
\end{pmatrix} ,
\label{20}
\end{eqnarray}
\noindent \\
where $r$ is the reflection coefficient of the ring with one lead, see Fig.~\ref{fig1}. 

Substituting Eq.~(\ref{20}) into Eq.~(\ref{15}), I get,  
 
\begin{eqnarray}
\hat G ^{(1)}_{} &=&
\begin{pmatrix} 
G ^{(1)}_{}  & 
0  \\
0  & 
G ^{(1)}_{} 
\end{pmatrix} ,
\label{21}
\end{eqnarray}
\noindent \\
where  $G ^{(1)}$ is given in Eq.~(\ref{01}).  

Near the time when the degenerate levels in the ring crosse the Fermi level in the leads, I use Eq.~(\ref{05}) and represent  
$  \hat G ^{(1)}\left( t_{1}; t_{2} \right) = \sum_{j=L , R}^{} \hat \Psi_{j}^{*}\left( t_{1} \right) \otimes \hat \Psi_{j}\left( t_{2} \right)  $ with 

\begin{eqnarray}
\hat \Psi_{L} &=& 
\begin{pmatrix} 
\Psi  \\ 
0
\end{pmatrix} ,
\quad \quad
\hat \Psi_{R} = 
\begin{pmatrix} 
0  \\ 
\Psi  
\end{pmatrix} ,
\label{18}
\end{eqnarray}
\ \\ \noindent
where the wave function $ \Psi$ is shown in Eq.~(\ref{06}). 

Thus, from the above equations we see that two electrons are emitted from  degenerate levels. 
Each electrons is emitted into a separate lead. 
This is why both wave packets emitted simultaneously are the same. 

%%%%%%%%%%%%%
%%%%%%%%%%%%%
\section{Conclusion}
\label{concl}

I analyzed the adiabatic emission of electrons on-demand  from two  orbitally degenerate quantum levels of non-interacting spinless electrons.   
As a paradigmatic example, I considered a one-dimensional ballistic ring with  one or two leads attached. 
The energy of the levels of the ring relative to the Fermi energy of the leads is changed using a uniform electric potential along the ring. 
As a convenient theoretic tool, I used an excess correlation function of electrons in  leads to predict the properties of single-electron wave packets emitted from a ring.

In a ring with one lead, under quite natural conditions, when electrons moving to the left and to the right are equally coupled to the lead, only one electron is emitted when the orbitally degenerate levels in the ring cross the Fermi energy in the lead. 
The other electron remains in the ring because its state is decoupled from the lead. 
When the degeneracy is lifted, say, with the help of the Aharonov-Bohm magnetic flux through the ring, both electrons are emitted at different times.  
I have analyzed in details how the emission of a second electron is  suppressed by a decrease in magnetic flux. 
First, the time difference between emissions is reduced to zero. 
In this case, two simultaneously emitted electrons have the same density profile, but different energies. 
With a further decrease in the magnetic flux, the width of one of the wave packets begins to decrease.  
A narrowing of the wave packet indicates a decrease in the effective coupling  between the corresponding level and the lead. 
As a result, the emission process for this level becomes non-adiabatic, which, in turn, increases the time difference between the emissions of two electrons. 
And finally, the emission of the second electron is infinitely delayed. 

When one additional lead is connected, the emission scenario may or may not change depending on the properties of the leads. 
If two symmetrically coupled leads are identical, only one electron is emitted from degenerate levels in the ring, it delocalizes between the two leads. 
To achieve two-electron emission, it is necessary to lift the degeneracy, for example, using a magnetic flux.
In contrast, if one of the leads has a bound state with the Fermi energy at its   end, both electrons are emitted without lifting the degeneracy, with each electron being injected into a separate lead. 
Therefore, if such a bound state can be created and destroyed with the help of some external influence, then the change in the number of electrons emitted from degenerate levels can be used as an indicator of the appearance of a bound state at the end of one of the leads. 

%%%%%%%%%%%%%
%%%%%%%%%%%%%
%\acknowledgments
\section*{Acknowledgments}

M.M. acknowledges the warm hospitality and support of Tel Aviv University, and support from the Ministry of Education and Science of Ukraine (project No. 0119U002565).

%%%%%%%%%%%%%
%%%%%%%%%%%%%
%\section*{References}

\end{document}